\documentstyle[twocolumn,prb,aps]{revtex}

\begin{document}
\draft
\preprint{Ames Laboratory-USDOE Preprint }
 
\title{Localization in Highly Anisotropic Systems}
 
\author{I. Zambetaki,$^{1}$ Qiming Li,$^{2}$ E. N. Economou,$^{1}$ and
 C.~M.~Soukoulis,$^{1,2}$ }
 
\address{$^1$Research Center of Crete and Department of Physics\\  P. O.
  Box 1527, 71110 Heraklion,
  Crete, Greece\\}
 
\address{$^2$Ames Laboratory and Department of Physics and Astronomy,\\
  Iowa State University, Ames,
  IA 50011\\}
\author{\parbox[t]{5.5in}{\small
 The localization behavior of the Anderson model with anisotropic hopping
  integral t for weakly coupled planes and
  weakly coupled chains is investigated both numerically with the transfer
 matrix method and analytically within the
 self-consistent theory of localization.  It is found that the mobility edge
  is independent of the propagating direction.  However, the correlation
 (localization) length in the extended (localized)
  side of the transition can be very different for the two directions.  The
  critical disorder W$_c$ is found to vary from t$^{1 \over 4}$ for weakly
 coupled
 planes to t$^{1 \over 2}$ for weakly coupled chains.
}}
 
\maketitle
 
\normalsize
 
 The problem of Anderson localization in anisotropic systems has attracted
 considerable
 attention\cite{bib1,bib2,bib3,bib4,bib5} in recent years, largely due to
 the fact that the high T$_c$
 superconductors are highly anisotropic. Transport in the normal state is
 metallic in
 the ab-plane but appears semiconductor-like in the c-axis\cite{bib6}.
  The nature of the
 c-axis transport in high T$_c$ materials  is still controversial and its
 understanding
 may have important consequences for the
 theories of the normal and the superconducting state. This paradoxical property
 has prompted the proposal\cite{bib7} that a high T$_c$ material in the normal
 state is
 actually an insulator, appearing
 metallic only because the inelastic length in the plane is less than the
 localization
 length.  It has also been argued\cite{bib2,bib6} that a negative d$\rho$/dT
 in the c direction alone may
 signify anisotropic localization, with a metal-insulator transition
 depending on
 the propagating direction, in direct contradiction to the predictions of
 the scaling
 theory of localization\cite{bib8}.  A recent diagrammatic
 calculation\cite{bib1}
  lent support to such a claim.
 Previous diagrammatic analysis\cite{bib9}, however, led to the conclusion
 that the scaling property in
 anisotropic systems remains the same as that of the isotropic systems with
 a simple
 substitution of the conductance by its geometric mean. Given the perturbative
 nature of all the previous work, it is important to carefully study the
 localization
 behavior of disordered anisotropic systems with reliable numerical
 techniques and
 to determine whether the scaling theory is valid for highly anisotropic
 systems\cite{bib8,bib9,bib10}.
 
 In the present work, we have systematically studied the localization
 properties of a three-dimensional disordered anisotropic system described
 by a tight-binding Hamiltonian  with random on-site energies and anisotropic
 hopping integrals. We calculate numerically both the conductance and the
 localization
 length of such systems with the transfer-matrix method\cite{bib8}.
 By doing finite-size scaling analysis on sufficiently
 large systems, we find the metal-insulator transition
 occurs at the same critical disorder $W_c$ in all directions. Remarkably, at
 criticality, the geometric mean of the ratio of the finite-size
 localization length to the width of the bar is independent of the anisotropy.
 This underlies the conformal invariance property at the Anderson transition.
 For systems with extremely weak plane-coupling, the critical disorder $W_c$
 is found
 to vary
 with anisotropy t as t$^{1\over 4}$, in good agreement with a self-consistent
 theory of localization.  However, the correlation lengths in the extended
 regime, as
  well as the localization lengths in the localized regime, differ tremendously
 in different directions.  As we will
  discuss later, this difference of the correlation lengths in the
  propagating directions may  explain
  the c-axis transport of the high-T$_c$ materials.  The localization
  behavior of weakly coupled
  disordered chains, which simulate the 1D-to-3D behavior, is also studied.
  The critical disorder W$_c$ is found to behave as
  t$^{1 \over 2}$.
 
  We study the following Hamiltonian for an anisotropic 3D disordered model
\begin{equation}
 H = \sum_n \epsilon_n |n >  < n| + \sum_{n,m} t_{nm} | n >  < m |
\end{equation}
  where the sites $\{n\}$ form a regular cubic lattice.  The hopping integrals
 are nonzero between  nearest-neighbor sites only and depend on directions,
 in general,
 $t_x \ne t_y \ne t_z$. We normalize all energies by the largest hopping
 integral.
 For systems with weak plane coupling, the hopping integrals are given by
 $\{1,1,t\}$.
 As a convention,  we have assigned the direction with the large and small
 hopping
 integral as the parallel ($\parallel$) and the perpendicular ($\perp$)
 direction,
 respectively.   Disorder is introduced by choosing randomly site energies
 $\epsilon_n$ within [-W/2, W/2].
 
 The metal or insulating nature of the Anderson Hamiltonian, Eq. (1), can be
 determined by investigating the scaling properties of either the
 conductance or the localization length of finite systems calculated
 from the transfer-matrix techniques\cite{bib8,bib11}. The conductance G of
 an M $\times$ M $\times$ M cube is calculated from the multi-channel Landauer
 formula\cite{bib12}
\begin{equation}
 G(M)=\frac {e^2} {h} Tr(T^+T)
\end{equation}
 where T is the transmission matrix. For the localization length calculation,
 one considers a bar of length N and cross-section M $\times$ M. One
 determines the largest localization length $\lambda_M$ as N
 $\to \infty$ from the smallest Lyapunov coefficient of the product of the
 random transfer matrices.
 From a plot of $\lambda_M$/M (or G) vs. M, one can determine the localization
 properties of the system.  For localized
  states (W $>$ W$_c$) the ratio $\lambda_M$/M (or G) is expected to fall with
  increasing M, while for extended
  states (W $<$ W$_c$) $\lambda_M$/M (or G) should rise instead.
  At the mobility edge trajectory (W=W$_c$),
  $\lambda_M$/M (or G) is independent of M and this behavior defines the
 Anderson
  transition point and the critical disorder W = W$_c$.
 In our calculations, we have used systems with M = 5 - 17 and N
  was at least 5000.  For the conductance calculations, M was up to 22.
 Due to the nonself-averaging
 nature of finite-size systems, an average over many
 random configurations  (up to 500 for the M = 20 case)
 must be taken to suppress the large fluctuations.
 
 In Fig. 1, we present our numerical results for the dimensionless
 conductance  $g=\frac {G} {e^2/h}$ and the scaled localization length
  $\lambda_M$/M  versus M at E=0  for the
  case of weakly coupled planes with coupling constant t = 0.1, for both the
  perpendicular (lower panels) and parallel (upper panels) directions.
 These results strongly suggest that for sufficiently large M, the critical
 disorder defined by
 an M-independent g and $\lambda_M$/M converges to the same value,  W$_c
 \simeq$8.4,
 for both propagating directions.  Notice that if only sizes M$\le$ 11 were
 used, which are appropriate in the isotropic case,  one would then have
 erroneously concluded  that W$^\parallel_c$ $>$ W$^\perp_c$.
  Another important point is that
 the value of $\lambda_M$/M at W = W$_c$, called $\Lambda_c$, is different in
 the two propagating directions.  We find that
  $\Lambda^\perp_c$ = t$\Lambda^\parallel_c$ for all the t's we have examined.
 The critical conductance in the two directions at t=0.1 is approximately
  g$^\perp_c \simeq$ 10$^{-6}$g$^\parallel_c$. This is a much stronger variation
 than the  g$^\perp \simeq$ t$^2$ g$^\parallel$
 predicted by the diagrammatic analysis\cite{bib9} of the anisotropic model.
 This may be an indication of the existence of high order corrections to the
 conductivity that is not simply proportional to the bare conductivity.
 This point needs further study.
 
  In order to extrapolate to infinite system size (M$\to \infty$),
  it is necessary to investigate the scaling behavior of $\lambda_M$/M.
  In the center of the band (E = 0) and close enough to the transition
 point, it turned out to
  be possible to establish a scaling function for both propagation
 directions within the accuracy
  of our numerical results.
  The scaling function f(x) behaves as 1/x in the x
 $\rightarrow$ 0 limit
 for extended states (while for localized states f(x) $\sim$ x).  For x
 $\rightarrow \infty$,
 f(x) approaches a constant value that depends on the propagating direction.
  For t = 0.1, we obtained $\Lambda_c^{\parallel}$ = 1.2 and
  $\Lambda^\perp_c$ = 0.12.
  This suggests that $\Lambda_c^\perp$ = t$\Lambda^\parallel_c$ for general t.
  We have, indeed, checked that this formula is correct for any t.
  Another important relation we were able to obtain is the
 geometric mean of the different
  critical values $\lambda_M$/M in the different directions is independent of
 the anisotropy t.
  We derived \begin{equation}(\Lambda_c^x \Lambda_c^y \Lambda_c^z)
 ^{1/3} = 0.6 \end{equation}
  In the case of weakly coupled planes, Eq. (3) becomes
 $((\Lambda_c^\parallel)^2
  \Lambda^\perp_c)^{1/3}
  \simeq$ 0.6.  This relation may have important consequences for the
  existence of conformal
  invariance at the critical point of the Anderson localization problem
 \cite{bib13,bib14}.
 
  In Fig. 2, we plot the localization length L$_c$ and  the correlation length
  $\xi$ as a function of W for
  E = 0 and t = 0.1 for the parallel and the perpendicular directions.
  Both L$_c$ and $\xi$
  diverge at the critical disorder W$_c$.  However, both $\xi$ and L$_c$
  differ substantially for the
  different propagating directions.  In particular, in the extended regime
  where W$<$W$_c$,
  $\xi^{\bot} >   \xi^{\|}$.  For example, at W=5, $\xi^{\bot}$=100, and
  $\xi^{\|} \simeq$1.5, in units of the lattice constant, which has been taken
 to be one.  In the
  localized regime where W$>$W$_c$, L$^{\|}_c$ $>$ L$^{\bot}_c$.  For example at
  W=10, L$^{\|}_c$=55, and L$^{\bot}_c \simeq$5. Our numerical results  
  approximately follow the theoretical prediction\cite{bib9} 
  $\xi^{\|}=t^2 \xi^{\bot}$ and $L_c^{\bot} = t L_c^{\|}$.
 However, the exponents are
 expected to be the same and this has been confirmed from our calculations. We
 find that $\nu^\bot=1.3\pm 0.1$ and $\nu^\|=1.3 \pm 0.3$,
 in agreement with each other and with $\nu=1.3 \pm 0.1$ for the
 isotropic system  within the numerical accuracy.
 
 The difference between $\xi^{\|}$ and $\xi^{\bot}$ is
  very important and can possibly explain the normal state
 transport properties\cite{bib6} of the high-T$_c$ materials.
 The correlation length $\xi$ measures the strength of the fluctuations of
 the wave functions in the extended regime.  For length
 scales larger than $\xi$, the wave function looks uniform, while for length
 scales smaller than $\xi$ the wave function has strong fluctuations.
  The relevant
 length scale is the inelastic mean free path $\ell_{in}$ which behaves as
 T$^{-p}$, with probably p=1/2.
 When $\ell_{in} < \xi$, a phenomenon called incipient localization takes place
 and conductivity is controlled by $\ell_{in}$.
 A convenient interpolation formula\cite{bib15}, valid in the conducting regime
  is given by
\begin{equation}
 \sigma_{loc}=\frac {e^2}{\hbar}(\frac {a}{\xi}+\frac {b}{\ell_{in}})
\end{equation}
 where a and b are constants of order unity\cite{bib15a}.
For high enough temperatures
  the conductivity is given by the regular metallic behavior,
 where $\sigma_{ph}=\omega_p^2\tau/4\pi$
 and the mean free time  $\tau \sim T^{-1}$. Because $\sigma_{ph}$ and
 $\sigma_{loc}$
  describe independent physical processes, we can add the corresponding
 resistivities.
 Therefore $\rho=\rho_{loc}+\rho_{ph}$, where $\rho=1/\sigma$.
  The experimental behavior of the high-T$_c$ materials can be
 understood if we assume that the
 high-T$_c$ oxides, instead of being insulators\cite{bib7},  are disordered
 anisotropic metals with a large anisotropic correlation length.
 At low T, the resistivity is dominated by $\rho_{loc}$. Once $\ell_{in}$
 becomes shorter than $\xi$ in the perpendicular direction
  (c-direction), $\xi^{\bot}$, Eq. (4) suggests
 that there is an sharp downturn in the perpendicular resistivity as the
 temperature increases. This trend will eventually stop at some T when the
 regular resistivity begin to dominate. Then $\rho^{\bot}$ will start
 increasing
 linearly with T, as in the regular metallic behavior.
 In the parallel direction,
  $\xi^{\|}$ is always smaller than the inelastic
  length and the transport in
  the plane remains metallic. The above analysis neglects correlations and
 dynamic disorder that could also affect substantially the  transport
 properties\cite{bib2}.
 
 The dependence of $W_c$ on the anisotropy t for the weak plane coupling
 has been also numerically calculated by the transfer matrix technique.
 The numerical values of $W_c$ can be fitted well
 with a single power law dependence W$_c$=15.4 t$^{1 \over 4}$, for $t<0.8$.
 This is clearly seen in Fig. 3.
 This dependence is in marked contrast to what one would expect based on
 reasonable heuristic argument\cite{bib16} and also predicted previously
 by more elaborate theories\cite{bib1,bib4} which give a much  weaker
 t dependence, W$_c \sim 1/\sqrt{\vert \ell n t\vert}$. The t dependence is
 also different from the results obtained  for the weakly coupled
 chains\cite{bib17}, which give W$_c \sim$t$^{1 \over 2}$. To
 understand the
 t$^{1\over 4}$ dependence of W$_c$, we examine the localization transition
 starting from the results of the diagrammatic analysis\cite{bib9}. The
  maximally crossed diagrams produced a
  correction\cite{bib4} to the zero-temperature configurationally average
 conductivity
  $\sigma_{io}$ (i=x,y,z) of
  the form given in Eq. (2.3a) of Ref. 4.
  An equivalent
 criterion in the tight-binding representation for the mobility edge is given
 by Eq. (2.5) of Ref. 4.
 For the anisotropic case we made the choice that the effective lattice constant
 $a_i$ is proportional to anisotropic mean free path $\ell_i$.
 $\ell_i \sim (<v_i^2 >)^{1/2}\tau = \big[ <v> 
  <\frac{v_i^2}{v}> \big]^{1/2}\tau$, i.e., we assume that $\tau$ is
  isotropic.
    Since $\sigma_{io}/a_i^2$ is independent
  of i, the localization criterion is satisfied simultaneously in all
 the directions and can be written as
\begin{equation}
  \frac{\pi\hbar\Omega\sigma_{io}}{2e^2 a_i^2} = \frac{1}{(2\pi)^3} \int
  d\vec{q}\frac{1}{\sum_{i=1}^3
  (2-2\cos q_i)} = G_{3D}^{is.} (E=6) ,
\end{equation}
  where  $\Omega =\prod_{i=1}^3 a_i$
  and 2G$_{3D}^{is.}$(E=6)=0.505462 is the 3D Green's function of the
 isotropic system at the band edge\cite{bib18}. Notice the same equation
 applies to the isotropic case except now both the conductivity
 and the mean free path enter as geometric means.
 
  In the limit of weak coupling t$\to$0, Eq. (5) yields
 $\tau \sim t^{-{1\over 2}}$ for weakly coupled planes and $\tau \sim
 t^{-1}$ for
 weakly coupled chains\cite{bib19},  and therefore  W$_c$
  ( taking into account that $\tau\sim$ W$^{-2}$ ) is proportional to
  t$^{1 \over 4}$ and  t$^{1 \over 2}$, respectively, in
  agreement with the transfer-matrix method results. We have systematically
 calculated\cite{bib19} W$_c$ vs t, for general t, within the CPA, by using
 Eq. (5).
 There is quantitative agreement between the CPA and the transfer matrix
 results.
   We feel the choice of the effective lattice constant a$_i$ being
  proportional to an anisotropic mean free path $\ell_i$ is the
  proper one and the good agreement with our numerical results further
  supports this.
 
  In summary, we have numerically studied, by the transfer-matrix method
  technique, the localization
  properties of weakly coupled planes.  We found that only one
  mobility edge exists for both
  propagation directions, i.e., the states in the planar direction become
  localized with exactly the
  same amount of disorder as the states in the perpendicular direction.
  However, the correlation length
  $\xi$ in the extended side of the transition, as well as the localization
  length in the localized side,
  can be very different for the two propagating directions.  This behavior of
  $\xi$ can possibly  explain the
  transport properties of high-T$_c$ materials.  The critical value of
  disorder W$_c$ is proportional to
  t$^{1 \over 4}$ for weakly coupled planes and is proportional to
 t$^{1\over 2}$ for
  weakly coupled chains.
  These results are found to be in satisfactory agreement with the
  predictions of the self-consistent theory of localization which incorporates
 the idea of length scale rescaling. However, the conductance in different
  directions does not satisfy the relation predicted by the diagrammatic
 analysis\cite{bib9}. Work to further understand the scaling properties
 of the conductance are in progress.
 
\acknowledgements
 We are grateful to A. G. Rojo for sending us his numerical data
 at the early stage of the project.
  Ames Laboratory is operated for the U. S. Department of Energy by Iowa
  State University
  under contract No. W-7405-ENG-82. This work was supported by the Director
  of Energy Research, Office of
  Basic Energy Sciences and NATO Grant No. CRG 940647. It was also
  supported by EU Grants.

\begin{figure}
 \caption{The conductance g (a and b) and $\lambda_M$/M (c and d) plotted as a
  function of M for E=0, t=0.1, and various values of disorder W, for both
  propagating directions.  The mobility edge is at W$_c \simeq$8.4 for both
  propagating directions.
  \label{1}}\end{figure}
 
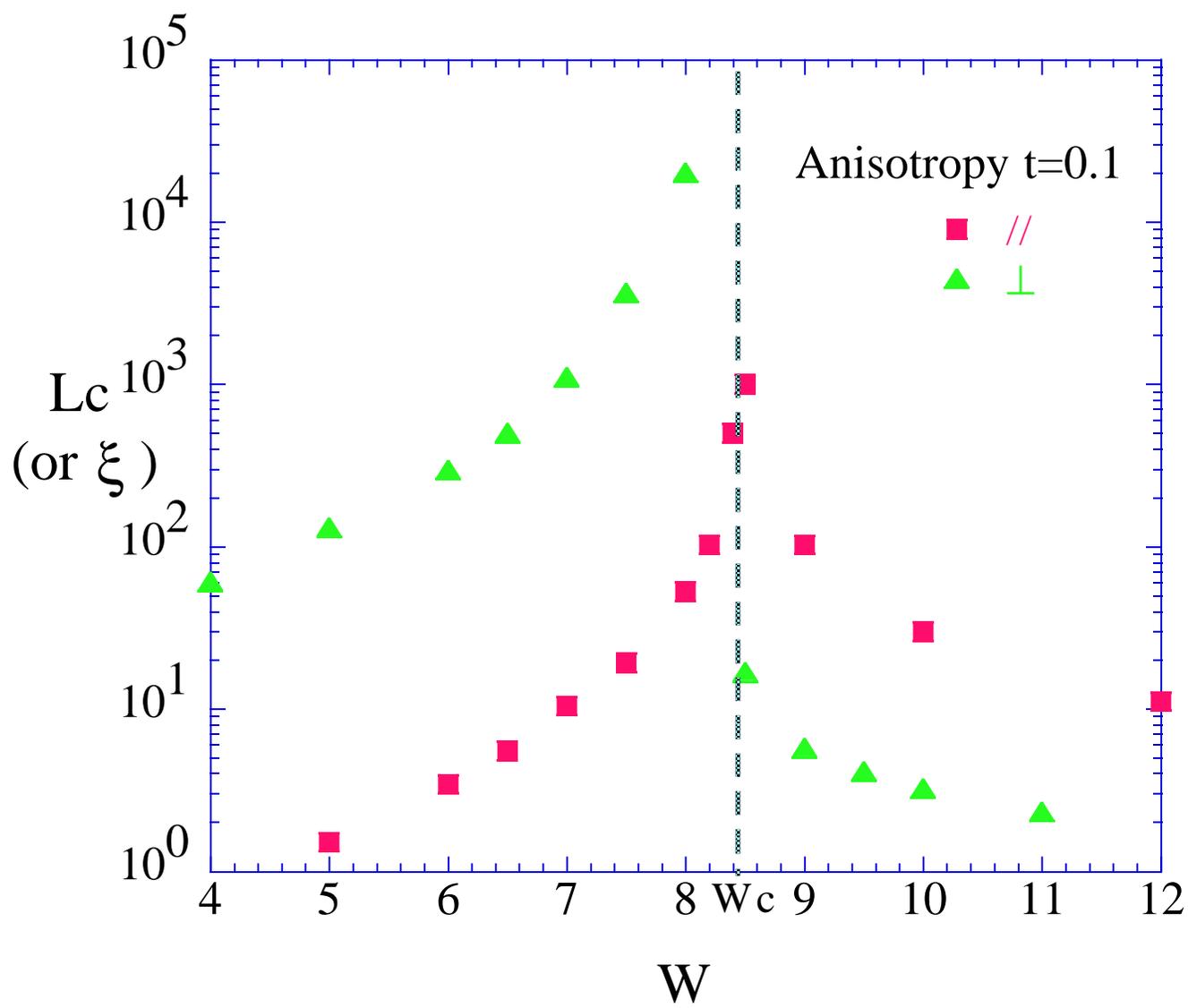
\begin{figure}\caption{Localization length L$_c$ or correlation length
  $\xi$, obtained by finite size scaling, as a
  function of disorder W for E=0 and t=0.1, for the parallel and
  perpendicular propagating directions.  Notice that for W$<$W$_c
 \simeq$8.4,
  $\xi^{\bot}> \xi^{\|}$, while for W$>$W$_c$, L$_c^{\|}$ $>$ L$_c^{\bot}$.
  \label{2}}\end{figure}
 
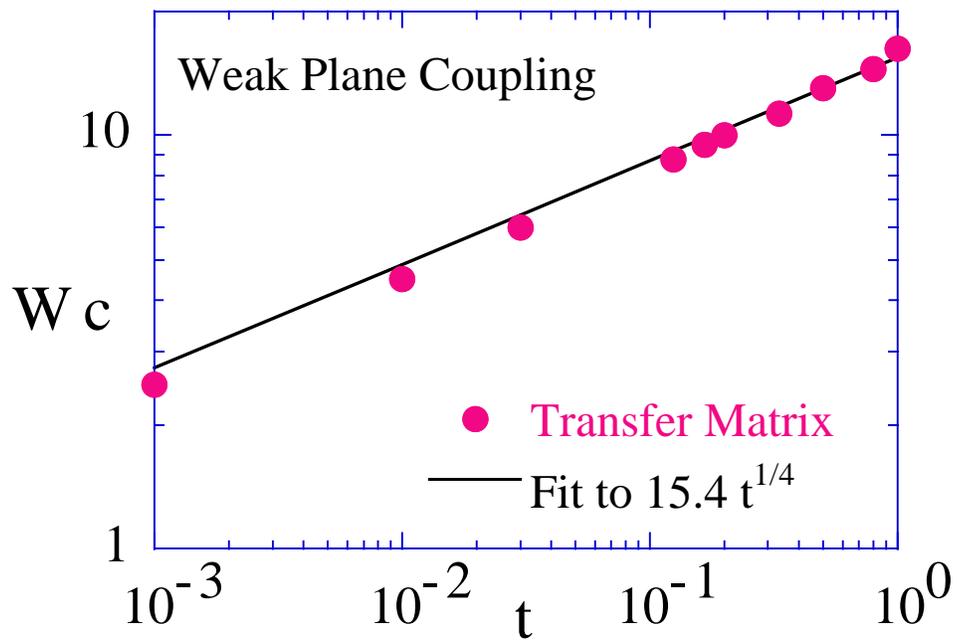
\begin{figure}
 \caption{The critical disorder W$_c$ for obtaining localized states at E=0
  vs the anisotropy t.
  \label{3}}\end{figure}
 

\begin{references}
\bibitem{bib15a}
Eq. (4) is strictly valid in the weak disorder limit. This can be seen 
by expanding g$\sim$ 1/(exp(2M/$\lambda_M$)-1) for weak disorder, where 
$\lambda_M >> M$. So g $\sim$ $\lambda_M/M$ and by using the 
data in Fig. 1(c-d), we obtain g $\sim$ b + aM/$\xi$, 
and therefore $\sigma \sim$ b/M
+ a/$\xi$. The coefficient b is approximately a constant of 
order $\Lambda_c$ away from the critical regime. Notice that $\Lambda_c^\|$ = 
1.2 and $\Lambda_c^\bot$ = 0.12 for t = 0.1.
\bibitem{bib16} To determine the critical disorder W$_c$ where the system
  crosses over from 1D-to-3D
  for the weakly coupled chains and from 2D-to-3D for the weakly coupled
  planes we consider a particle in
  a particular chain or plane, respectively.  With probability $\frac
 {1}{2(3-d)}$
 the particle will hop to another
  chain or plane after a time and therefore distance proportional to 1/t,
  where t is the hopping integral.
  The localization length L$_c$ in the 1D (2D) case is proportional to
  $\frac {1} {W^2} (exp(\frac {\alpha}{W^2}))$,
  respectively.  So the condition L$_c \simeq$1/t, determines the critical
  disorder W$_c$.  For the
  1D-to-3D, we have 1/W$_c^2 \simeq 1/t$, and therefore W$_c \sim t^{1
 \over 2}$.
  For the 2D-to-3D we have
  $exp ( \frac {\alpha} {W_c^2}) \simeq 1/t $ and therefore W$_c \sim
 1/\sqrt{\vert \ell nt\vert}$.
\end{references}
\end{document}